\documentclass[conference]{IEEEtran}
\IEEEoverridecommandlockouts
\usepackage{cite}
\usepackage{amsmath,amssymb,amsfonts}
\usepackage{algorithmic}
\usepackage{graphicx}
\usepackage{textcomp}
\usepackage{xcolor}
\usepackage{xspace}
\usepackage{subcaption} 
\usepackage{multirow}
\usepackage{booktabs}
\usepackage{makecell}
\usepackage{varwidth}
\usepackage{hyperref}
\usepackage{enumitem}

\def\BibTeX{{\rm B\kern-.05em{\sc i\kern-.025em b}\kern-.08em
    T\kern-.1667em\lower.7ex\hbox{E}\kern-.125emX}}

\newcommand{\spgemm}{SpGEMM\xspace}

\begin{document}

\title{Extreme-scale many-against-many protein similarity search\\
}

\author{
  \IEEEauthorblockN{
    Oguz Selvitopi\IEEEauthorrefmark{1},
    Saliya Ekanayake\IEEEauthorrefmark{2},
    Giulia Guidi\IEEEauthorrefmark{3},
    Muaaz G. Awan\IEEEauthorrefmark{4},
    Georgios A. Pavlopoulos\IEEEauthorrefmark{5},
    Ariful Azad\IEEEauthorrefmark{6}, \\
    Nikos Kyrpides\IEEEauthorrefmark{7},
    Leonid Oliker\IEEEauthorrefmark{1}, 
    Katherine Yelick\IEEEauthorrefmark{3}\IEEEauthorrefmark{1},
    Ayd{\i}n Bulu\c{c}\IEEEauthorrefmark{1}\IEEEauthorrefmark{3}}
  \vspace{0.5em}
  \IEEEauthorblockA{
    \IEEEauthorrefmark{1}\textit{Applied Mathematics \& Computational Research Division, Lawrence Berkeley National Laboratory, USA} \\
    \IEEEauthorrefmark{2}\textit{Microsoft Corporation, USA} \\
    \IEEEauthorrefmark{3}\textit{University of California, Berkeley, USA} \\
    \IEEEauthorrefmark{4}\textit{NERSC, Lawrence Berkeley National Laboratory, USA} \\
    \IEEEauthorrefmark{5}\textit{Institute for Fundamental Biomedical Research, BSRC ``Alexander Fleming'', 34 Fleming Street, 16672, Vari, Greece} \\
    \IEEEauthorrefmark{6}\textit{Indiana University, USA} \\
    \IEEEauthorrefmark{7}\textit{Joint Genome Institute, Lawrence Berkeley National Laboratory, USA} \\
    roselvitopi@lbl.gov}   
  }

\maketitle

\begin{abstract}
Similarity search is one of the most fundamental computations that are regularly performed on ever-increasing protein datasets. Scalability is of paramount importance for uncovering novel phenomena that occur at very large scales. We unleash the power of over 20,000 GPUs on the Summit system to perform all-vs-all protein similarity search on one of the largest publicly available datasets with 405 million proteins, in less than 3.5 hours, cutting the time-to-solution for many use cases from weeks. The variability of protein sequence lengths, as well as the sparsity of the space of pairwise comparisons, make this a challenging problem in distributed memory.  Due to the need to construct and maintain a data structure holding indices to all other sequences, this application has a huge memory footprint that makes it hard to scale the problem sizes. We overcome this memory limitation by innovative matrix-based blocking techniques, without introducing additional load imbalance. 

\end{abstract}


\section{Justification for ACM Gordon Bell Prize}
We unleash the power of over 20,000 GPUs to perform many-against-many protein similarity search on one of the largest publicly available datasets with 405 million proteins in 3.4 hours with an unprecedented rate of 691 million alignments per second, cutting the time-to-solution for many use cases from weeks.

\section{Performance Attributes}
\begin{table}[h]
  \begin{center}
    \scalebox{1.00} {
      \begin{tabular}{l l} 
        \toprule
        \textbf{Performance Attribute} & \textbf{Value} \\
        \midrule
        Category of achievement & \makecell[l]{Time to solution, alignments per seconds, \\ cell updates per second (CUPs)} \\
        \midrule
        Type of method used & N/A \\
        \midrule
        \makecell[l]{Results reported on the \\ basis of} & \makecell[l]{Whole application for time to solution \\ and alignments per second. \\ Kernel time for cell updates per second} \\
        \midrule
         Precision reported & Integer \\
        \midrule
        System scale & \makecell[l]{3364 nodes \\ (141,288 CPU cores and 20,184 GPUs)} \\
        \midrule
        Measurement mechanism & Timers \\
        \bottomrule
      \end{tabular}
      }
  \end{center}
  \label{tb:perf-attr}
\end{table}

\section{Overview of the Problem}
Comparative genomics studies the evolutionary and biological relationships between different organisms by exploiting similarities over the genome sequences.
A common task, for example, is to find out the functional or taxonomic contents of the samples collected from an environment often by querying the collected sequences against an established reference database.
%
%
The importance of enabling and building of fast computational infrastructure for comparative genomics becomes more critical as more and more genomes are sequenced.

Our work addresses the computational challenges posed by searching similarities between two sets of proteins in the sequence domain.
%
The use cases of this task in computational biology are numerous and include functional annotation~\cite{Glover2019}, gene localization and studying protein evolution~\cite{Caetano-Anolles2003}.
In metagenomics the DNA sequences collected from the environment enable the study of a diverse microbial genome pool that is often missed by the cultivation-based methods.
Such samples contain millions of protein sequences~\cite{Godzik2011} and a major component of many biological workflows is to find out the existing genes by aligning them against a reference database.
With the sequencing costs dropping and the technology becoming more available, the bottlenecks in metagenomics research are gradually shifting towards computation and storage~\cite{Scholz2012, Prakash2012}.
%

We focus on the problem of aligning a set of sequences against another set of sequences.
This problem often occurs within the context of identifying sequences in one set (set of query sequences) by using another set of sequences whose functions are already known (set of reference sequences).
Another context is to find the similar sequences in a given set by clustering them.
In this variant, a many-against-many search is performed over a set of sequences to find the similar sequences in the set (often followed by clustering of sequences).
This variant can also be seen as aligning the given set against itself where the query and the reference set is the same.
We refer to this as many-against-many protein similarity search and focus on this search problem in our work.

Our work demonstrates that HPC is a viable fast alternative in enabling tree-of-life scale metagenomics research.
Harnessing the power of accelerators which are well suited to SIMD-type parallelism that are required by the alignment operations in the search, we develop novel parallel algorithms and optimization techniques that are able to simultaneously utilize all resources on the nodes and attain high performance.
The key points in our approach for addressing the described challenges can be summarized as follows:
\begin{itemize}[topsep=0pt,itemsep=-1ex,partopsep=1ex,parsep=1ex,leftmargin=*]
    \item The immense computational resources required by the large-scale search operations are met by distributed utilization of accelerators. Compute-intensive alignment operations form the main computational bottleneck and popular tools in this field~\cite{Buchfink2021, Steinegger2017, Doring2008} make use of SIMD parallelism on the CPUs with vector instructions but they do not utilize accelerators which are more suited to these types of operations.
    \item We take advantage of the heterogeneous architecture of the nodes and  hide the overhead of memory-bound distributed overlap detection component of the search by performing them on the CPUs simultaneously with the alignment operations on the accelerators.
    \item To avoid IO during the search--which are often the method of choice with distributed tools in this domain when the scale of the search starts to become infeasible--we develop a distributed 2D Blocked Sparse SUMMA algorithm that performs the search incrementally and hence can effectively control the maximum amount of memory required by the entire search. In this way, our approach only uses IO at the beginning and at the end, which are both done in parallel and constitutes at most 3\% of the entire search time.
    \item By relying on custom load balancing techniques and distributed sparse matrices as the founding structures whose parallel performance is well-studied in numerical linear algebra, we obtain good scalability by attaining more than 75\% strong-scaling and 80\% weak-scaling parallel efficiency.
\end{itemize}

The biggest reported protein sequence similarity search on a supercomputer system to the best of our knowledge was in 2021 by DIAMOND~\cite{Buchfink2021}.
This search involved querying 281 million sequences against 39 million sequences on 520 nodes of the Cobra supercomputer at the Max Planck Society and took 5.42 hours by performing a total of 23.0 billion pairwise alignments in the very sensitive mode (1.2 million alignments per second).
With our search tool PASITS (Protein Alignment via Sparse Matrices), we significantly improve this by performing a search of 405 million sequences against 405 million sequences on 3364 compute nodes of the Summit supercomputer at Oak Ridge Leadership Computing Facility
Our search took 3.44 hours by performing a total of 8.6 trillion pairwise alignments at a rate of 690.6 million alignments per second.
Overall, we increase the scale of the solved problem by an order of magnitude (15.0x) and improve the performed alignments per second by more than two orders of magnitude (575.5x).

\section{Current State of the Art}
There are many protein similarity search tools in the literature and each of them has different search techniques that are refined over the years.
Among the more popular of these tools are BLASTP~\cite{Altschul1997}, MMSeqs2~\cite{Steinegger2017}, LAST~\cite{Kielbasa2011}, DIAMOND~\cite{Buchfink2015}, and USearch~\cite{Edgar2010}.
%
%
%
In terms of parallelism, almost all of the mentioned tools support some of parallelism with varying degrees of efficiencies.
The libraries such as DIAMOND, LAST, and MMSeqs2 have great support for on-node parallelism: they can take advantage of vector instructions, can use multiple cores, have cache-friendly algorithms within them, etc.
Some of these tools such as DIAMOND, MMSeqs2, or mpiBLAST~\cite{Darling2003} also run in a distributed setting.
%
%
Some of the main shortcomings of these tools' distributed-memory parallelization can be summarized as follows:
\begin{itemize}[topsep=0pt,itemsep=-1ex,partopsep=1ex,parsep=1ex,leftmargin=*]
    \item In LAST and MMSeqs2, the index data structures for at least one set of the sequences (queries or targets) are replicated on each compute node before the search phase, which limits the largest problems that can be solved. In DIAMOND, they are written as partitioned chunks into disk, which severely increases the pressure on the file system. 
    \item The existing software do not have a global view of these replicas/chunks and the parameters are set per replica, which results in changing sensitivity with increased parallelism or memory constraints. For example, the DIAMOND guide states that ``this [block size] parameter affects the algorithm and results will not be completely identical for different values of the block size". By contrast, the PASTIS algorithm give identical results irrespective of the amount of parallelism utilized and the blocking size chosen.
    \item These search tools do not support GPUs. The GPUs harbor a much higher level of SIMD parallelism which are perfect fits for the pairwise alignments usually utilized in the search, which greatly enjoy this type of parallelism.
\end{itemize}

Among the protein search tools that have distributed-memory support, we further examine MMSeqs2 and DIAMOND as these two tools are the current state-of-the-art in distributed protein similarity search.

MMSeqs2~\cite{Steinegger2017} uses hybrid MPI/OpenMP for distributed-memory parallelism and has support for SSE and AVX2 vector instruction sets.
There are two modes of parallelism provided according to whether the reference or the query sequence set is distributed among the parallel processes.
In the first, the reference sequence set is divided into chunks and distributed among the parallel nodes.
In this mode, each process searches \emph{all query sequences} against its chunk of the reference.
In the second mode, the query set is divided into chunks and distributed among the parallel nodes.
In this mode, each process searches its chunk of queries against \emph{all reference sequences}.
%
%
In our earlier CPU-based PASTIS work, we found our approach to be more scalable than MMSeqs2~\cite{Selvitopi2020} and MMSeqs2 suffering from high IO overheads.

The distributed-memory parallelism in DIAMOND~\cite{Buchfink2021} is geared more towards providing capability to run on commodity clusters, i.e., cloud computing.
In this regard, it avoids using MPI that may not be found on such clusters and instead heavily relies on IO operations supported by POSIX-compliant parallel file systems.
DIAMOND divides both the reference and the query sequence sets into chunks and an element that is in the cartesian product of these two sets of chunks is referred to as a work package.
These packages are processed in parallel by worker processes.
This workflow makes a distinction between the parallel shared file system among nodes and the disks local to nodes.
Once the processing of a query chunk against all reference chunks is complete, the final worker process joins the results and writes them to an output file.
These choices may have serious performance implications for HPC systems.
Nevertheless, the focus of distributed-memory parallelism in DIAMOND is the capability to also run on commodity clusters and fault tolerance, rather than high performance.
%
%




\begin{figure}
	\centering
	\includegraphics[width=0.47\textwidth]{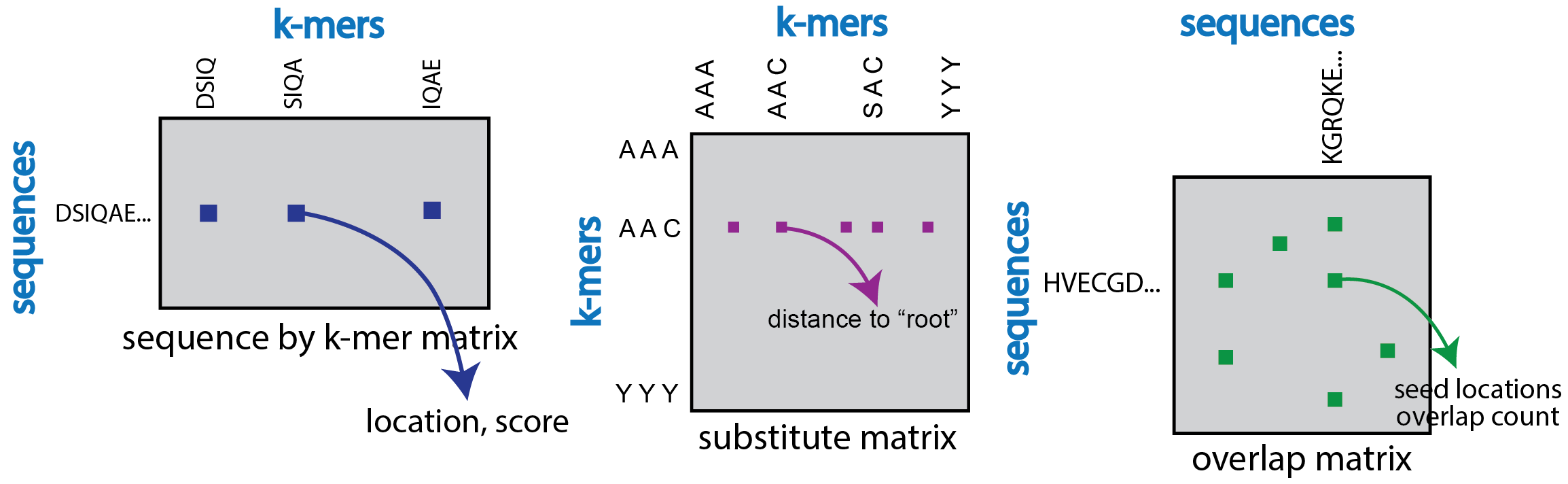}
	\caption{Examples of various sparse matrices used in PASTIS. The types of the elements in each matrix are different and a sparse matrix can utilize different element types according to the options provided (such as alignment type).}
	\label{fig:spmats}
	\vspace{-1.5em}
\end{figure}

\section{Protein similarity search pipeline}
Our approach for the protein similarity search problem consists of three main components: (i) discovery of candidate pairwise sequences which may harbor a certain degree of homology, (ii) batch alignment of the discovered candidate sequences, and (iii) forming the protein similarity graph from the information obtained in the alignment.
%
%
In the discovery of the candidate pairwise sequences, PASTIS has the option to introduce substitute $k$-mers that are $m$-nearest neighbors of a $k$-mer or plugging in a reduced alphabet~\cite{Murphy2000}, both of which can enhance the sensitivity.
It can make use of different alignment libraries and algorithms within them and can seamlessly integrate common sequence alignment metrics such as average nucleotide identity and coverage in the formation of the similarity graph.
These options enable PASTIS to reach out different regions of the overall search space and increase the effectiveness of the search.

The basic information storage and manipulation medium PASTIS is sparse matrices.
They are used to represent different types of information required throughout the search.
For instance, $k$-mer information in sequences are captured in a sparse matrix whose rows and columns respectively correspond to sequences and $k$-mers and a nonzero element in this matrix indicates the existence of a specific $k$-mer in a specific sequence.
Apart from being one of the most well-studied and optimized structures in parallel linear algebra, sparse matrices provide flexibility in the sense that any arbitrary information can be encapsulated within these elements, such as location, score, etc.
Figure~\ref{fig:spmats} illustrates some of the sparse matrices utilized in PASTIS and it can be seen  that these matrices easily allow to store and manipulate any necessary information required by our similarity search pipeline.
%

\begin{figure}[t]
	\centering
	\includegraphics[width=0.50\textwidth]{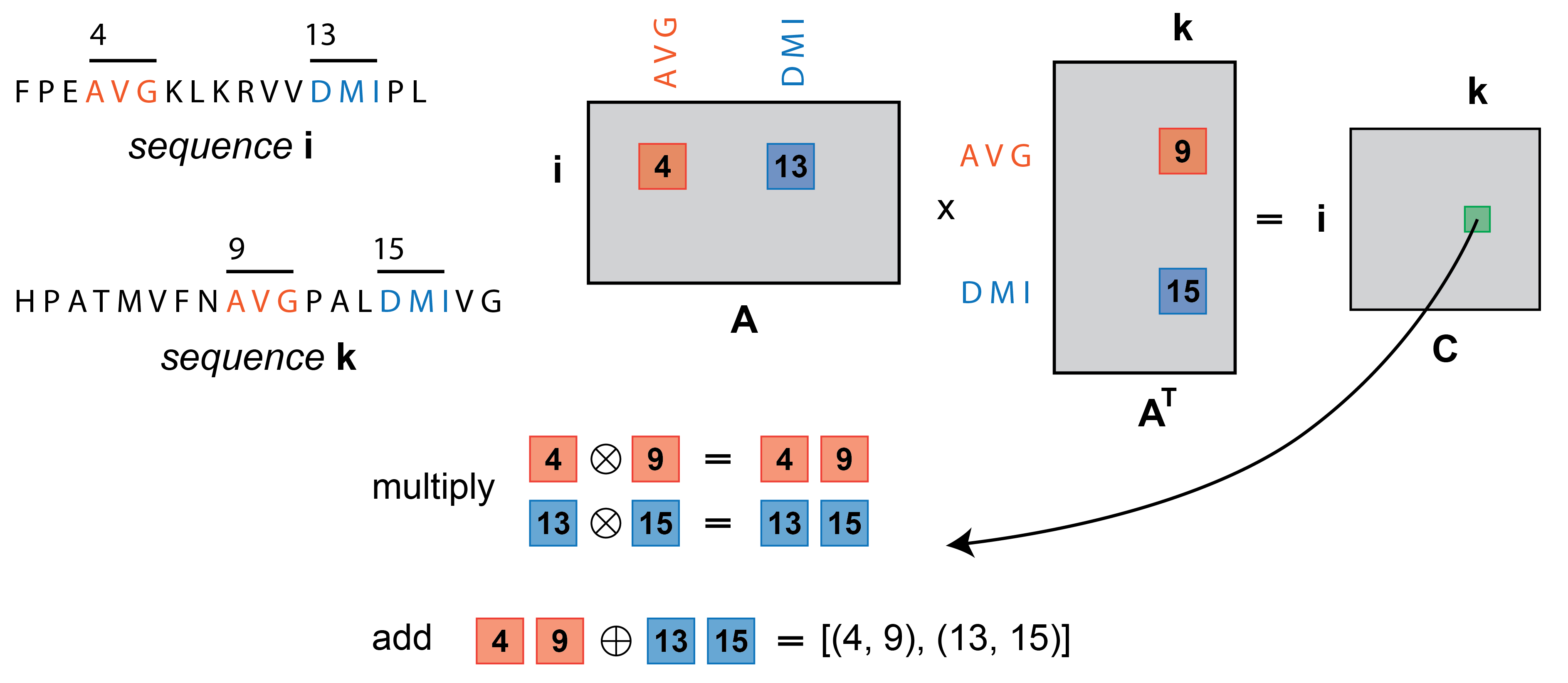}
	\caption{Semiring algebra allows PASTIS to express computations in similarity search through sparse operations. Here illustrated a simple example to discover a candidate pair for alignment.}
	\label{fig:semiring}
	\vspace{-1.5em}
\end{figure}

Although sparse matrices are very common and widely used in the field of linear algebra, their utilization and importance have recently started to gain momentum in graph computations thanks to the GraphBLAS standardization efforts~\cite{Kepner2016}.
The basic motivation is to express graph computations in the language of linear algebra and by doing so utilize decades of algorithmic and optimization work in sparse linear algebra within the graph computation frameworks.
Different from the matrix operations expressed in linear algebra, the operations on graphs usually require different operators to perform computations on sparse matrices.
For example in PASTIS, the discovery of candidate pairwise sequences is expressed through an overloaded sparse matrix sparse matrix ``multiplication'', in which the elements involved in this operation are custom data types and the conventional ``multiply-add'' operation is overloaded with custom operators, which are known as \emph{semirings}.
Semiring algebra allows to express graph operations through operations on sparse matrices and we utilize various semirings to enable different types of alignments (Figure~\ref{fig:semiring}).

\subsection{Software stack and parallelism}
\begin{figure}[t]
	\centering
	\includegraphics[width=0.37\textwidth]{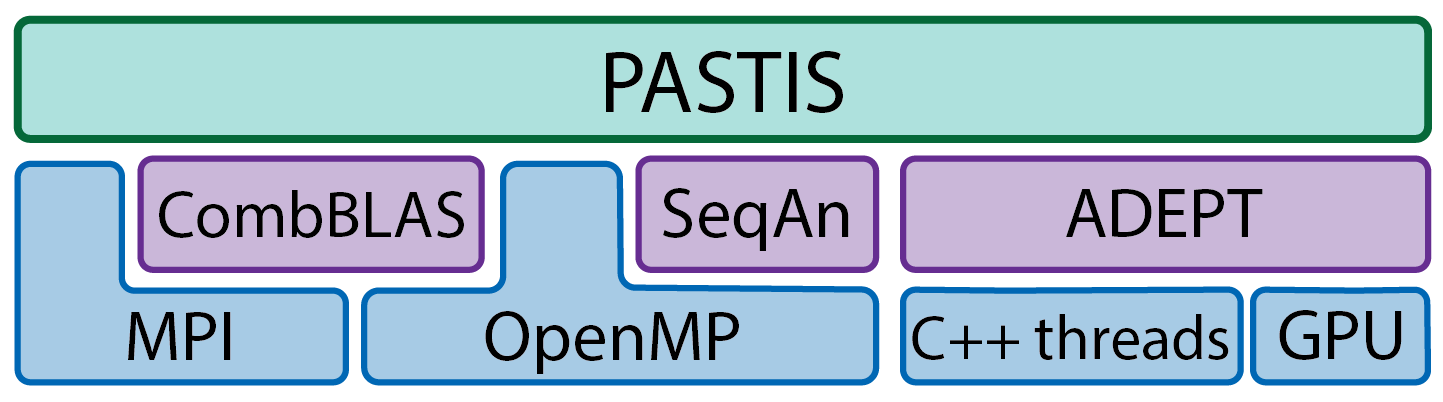}
	\caption{PASTIS utilizes different libraries and is able to make efficient use of both CPU and GPU resources found on a node.}
	\label{fig:software-stack}
	\vspace{-1.5em}
\end{figure}

Our protein similarity search pipeline utilizes several libraries and orchestrates them in a distributed setting.
For distributed sparse matrices and computations on them, it relies on CombBLAS~\cite{Azad2022} -- a distributed-memory parallel graph library that is based on arbitrary user-defined semirings on sparse matrices and vectors.
For parallel alignment, it utilizes SeqAn C++ library~\cite{Doring2008} and ADEPT~\cite{Awan2020}. 
Among these libraries, CombBLAS supports MPI/OpenMP hybrid parallelism, SeqAn supports node-level shared-memory parallelism with vectorization, and ADEPT supports node-level many-core parallelism
%
Apart from those PASTIS itself directly makes use of MPI/OpenMP hybrid parallelism.
The software stack of PASTIS is illustrated in Figure~\ref{fig:software-stack}.

An important design choice in our approach is to separate the parallelism level used for alignment and other components.
For alignment, we deliberately prefer on-node libraries that are able to exploit different aspects of parallelism found on the node such as threads, CPU vector instructions, or fine-grained parallelism on GPUs.
These on-node alignment libraries are handled in PASTIS through distributed sparse matrix computations for which there already exist fast and optimized data structures and algorithms.
The key to high performance, as we demonstrate in our work, is the good orchestration of on-node and node-level parallelism with techniques that are able to overcome the performance bottlenecks.

CombBLAS~\cite{Azad2022} uses a 2D decomposition for distributed sparse matrices in which the the matrices are partitioned into rectangular blocks.
It uses a square process grid with the requirement of number of processes to be a perfect square number.
It supports compressed sparse column and doubly-compressed sparse column sparse matrix storage formats~\cite{Buluc08} and contains fast and state-of-the-art algorithms for complicated operations like \spgemm, being able to run such operations efficiently both on-node level~\cite{Nagasaka2018} and on massive scale~\cite{Hussain2021}.
%

ADEPT~\cite{Awan2020} is a GPU accelerated sequence alignment library that supports both DNA and protein sequence alignments.
It uses a combination of inter- and intra-task parallelism approach to realize the full Smith-Waterman sequence alignment on GPUs.
ADEPT derives its performance from efficient use of GPU’s memory hierarchy and exploiting fast register-to-register data transfers for inter-cell communications while computing the dynamic programming matrix.
ADEPT has CUDA, HIP, and SYCL ports being able to utilize NVIDIA, AMD, and Intel GPUs, respectively. 
ADEPT’s driver class works as an interface between the calling application and the GPU kernels, the driver detects all the available GPUs on a node and distributes alignments across all the available GPUs.
A unique C++ thread handles data packing and transfers (to and from GPU) for each GPU. 



\begin{figure*}
	\centering
	\includegraphics[width=.93\textwidth]{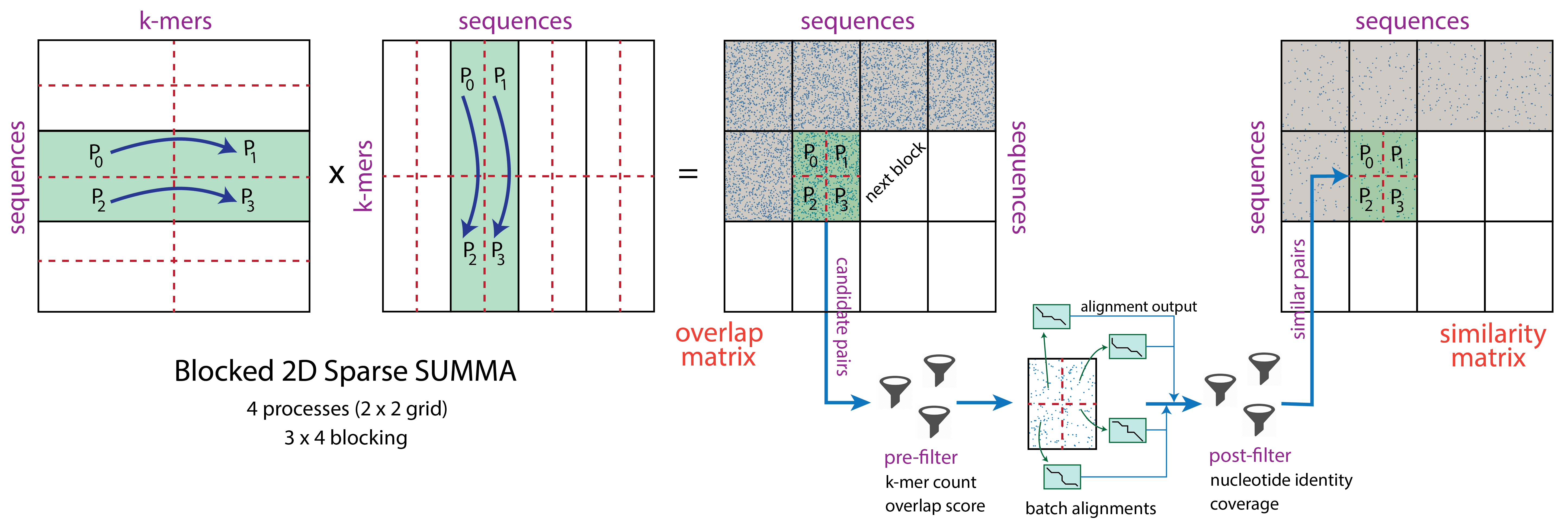}
	\caption{Discovery of the candidate alignments via Blocked 2D Sparse SUMMA and incremental similarity search. 
	}
	\label{fig:blocked-summa}
	\vspace{-1em}
\end{figure*}

\subsection{Performance characteristics}
\noindent \textbf{Computational patterns.}
The two basic types of computations performed in our protein similarity search pipeline are the sparse computations and edit distance computations required in the alignment.
The former is memory-bound having low computational intensity and high memory footprint with irregular access patterns while the latter is compute-bound having high computational intensity and a uniform  pattern in computing the edit distance matrices, which are small and dense.
Taking into account the fact that most alignment libraries are able to achieve high performance through SSE and AVX vector instruction sets (as is the case in SeqAn utilized by PASTIS) and these are not supported by IBM PowerPC processors on Summit, we dedicate GPU resources solely for alignment and utilize CPU resources for memory-intensive sparse computations.
Considering the mentioned characteristics of these computations, the accelerators are more suited for alignment than for sparse computations. 
For this reason, we rely on ADEPT (see Figure~\ref{fig:software-stack}) for alignment, which has on-node multiple accelerator support.
This library also uses resources on the CPU but they constitute a small percentage of the alignment.

\noindent \textbf{Memory requirements.}
Many-against-many protein similarity search requires huge amount of memory and the existing libraries in this area have various techniques to deal with this issue ranging from writing intermediate files to disks to performing the search in stages.
For a modest dataset containing 20 million sequences, one usually needs to store hundreds of billions candidate alignments and need to perform tens of billions of alignments.
The memory required by such a relatively small-scale search can quickly exceed the amount of memory found on a node.
Moreover, one also needs additional data structures to efficiently perform the search.
For example, the method to discover candidate alignments in our approach uses a parallel \spgemm, which usually needs much more intermediate memory than the actual storage required by the found candidates.
This factor, the average amount of intermediate results computed and stored per output element, is called the \emph{compression factor}, and even with a modest value between 1 and 10 that are often seen in genomics datasets, it is clear that memory management must be given special attention in many-against-many search.
Finally, the number of candidate pairs that need to be stored and aligned grows \emph{quadratically} with the number of sequences in the search, which makes the similarity search over huge datasets even more challenging in terms of memory requirements.

\noindent \textbf{I/O and communication.}
PASTIS uses parallel MPI I/O for input and output files.
The input to PASTIS is a file in FASTA format (a very common file format in bioinformatics to represent nucleotide and protein sequences) and the output is the similarity graph in triplets whose entries indicate two sequences and the similarity between them.
The output file is typically larger than the input file.
The communication in PASTIS can be categorized into two as the communication required for the sequences and for sparse computations.
The overhead of the former is effectively hidden by performing it in a non-blocking manner till they are required, which is when the pairwise alignments are to be performed.
The communication (and computation) required by the most sparse computations can also be hidden given that the nodes have accelerator support.
We investigate this issue in Section~\ref{sec:pb}.
Compared to memory and computational issues described so far, I/O and communication bottlenecks usually constitute less of a problem in our approach in performing many-against-many protein similarity search.
Typically, IO takes no more than 3\% of the overall execution time in PASTIS.

\section{Innovations Realized}
\label{sec:innov}
We address the computational challenges posed by the distributed protein similarity search mainly with three novel techniques.
The main performance bottleneck, huge memory requirement of the search, is addressed via proposing a blocked variant of the 2D Sparse SUMMA utilized in the distributed formation of the overlap matrix (Section~\ref{sec:blsumma}).
By relying on the observation that the overlap matrix is symmetric (similarity graph is undirected), we propose techniques to avoid significant amount of sparse computations and two different load-balancing schemes that exhibit different behavior based on the blocking factors and are able to achieve good computational load balance (Section~\ref{sec:lb}).
%
%
We then describe a technique that hides the overhead of memory-bound sparse computations as well as certain communication operations (Section~\ref{sec:pb}).
We validate the proposed innovations on small-scale datasets containing a few tens of millions sequences.
All experiments are conducted on the Summit system (see Section~\ref{sec:perf-res} for specs).
For all the experiments reported, we use 1 MPI task per node and utilize all 42 cores and 6 GPUs on each node.
%

\subsection{Blocked 2D Sparse SUMMA}
\label{sec:blsumma}
In our approach, the candidate sequences are discovered through a parallel \spgemm which produces an overlap matrix that contains pairs of sequences to be aligned.
The memory required by candidate pairs are huge and the motivation for blocked formation of the candidate pairs rests on the observation that only a fraction of them are actually similar.
Using the information available before and after the alignment (common number of $k$-mers, nucleotide identity, coverage, etc.), typically only less than 5\% of the candidate pairs end up in the final similarity graph.
Therefore, incremental similarity search can greatly reduce the memory used.

The algorithm for parallel \spgemm of form $C=AB$ used for the computation of the overlap matrix is the 2D Sparse SUMMA algorithm~\cite{Buluc2012}.
For our analyses, we assume the matrices are square having a dimension of $n$ for rows/columns and the elements of the sparse matrices are distributed uniformly.
Given $p$ parallel processes, this algorithm proceeds in $\sqrt{p}$ stages in which certain sub-matrices of the input matrices are broadcast and partial results for the output matrix are computed.
Assuming that the collective broadcasts use a tree algorithm~\cite{Chan2007}, its communication cost is given by
\[
    2 \alpha \sqrt{p} \log \sqrt{p} + 2 \beta s \sqrt{p} \log \sqrt{p},
\]
where $s$ is the number of nonzeros in a sparse sub-matrix of dimensions $n/\sqrt{p} \times n/\sqrt{p}$, $\alpha$ is the message startup time and $\beta$ is the per-word transfer time.

We generalize the 2D Sparse SUMMA algorithm with arbitrary blocking factors in order to form the output matrix in blocks.
We form the output matrix in \spgemm in $br \times bc$ blocks, where $br$ and $bc$ are respectively row and column blocking factors.
In the computation of the output block $C(r,c)$ required are the row stripe $A(r,*)$ and the column stripe $B(*,c)$.
Originally, the input matrices are distributed among $p$ processes in a $\sqrt{p} \times \sqrt{p}$ grid.
Therefore, to be able to compute the output matrix in blocks, $A$ must be split into $br$ row stripes and $B$ must be split into $bc$ column stripes.
Each of these row and column stripes must be distributed among $\sqrt{p} \times \sqrt{p}$ process grid.
The left of Figure~\ref{fig:blocked-summa} displays how the input matrices are distributed among four processes organized into $2 \times 2$ grid for a $3 \times 4$ blocking and the sub-matrices used in the computation of $C(1,1)$.

\noindent \textbf{Communication costs.} 
Compared against the plain SUMMA, the blocked variant increases the communication overhead as the input matrices need to be broadcast multiple times.
As mentioned earlier, the memory requirement is one of the main bottlenecks in the similarity search.
In addition, as will be described in Section~\ref{sec:pb}, the overhead of the output block computations can be hidden to a large degree, i.e., overheads of both the broadcasts and local sparse computations.
Nevertheless, the increase in the communication cost might be prohibitive when there are a lot of blocks and if its overhead cannot be hidden.
The overall communication cost of blocked variant is given by
\[
    2 \alpha (br \cdot bc) \sqrt{p} \log \sqrt{p} + \beta s (br + bc) \sqrt{p} \log \sqrt{p}.
\]
%

From similarity search perspective, the advantages and disadvantages of blocked 2D Sparse SUMMA are as follows:
\begin{itemize}
    \item Enables similarity search over huge datasets and reduces the consumed memory required by many-against-many sequence alignment.
    \item Opens the path for different types of optimizations and hence able to further reduce the overall time required by the search significantly. Up to 30\% reduction in the overall runtime can be obtained with the techniques described in Section~\ref{sec:lb}~and~\ref{sec:pb}, which complement the blocked formation.
    \item It increases the time required to discover candidate alignments due to increased communication and split sparse computations.
\end{itemize}

\begin{figure}
	\centering
	\includegraphics[width=0.3\textwidth]{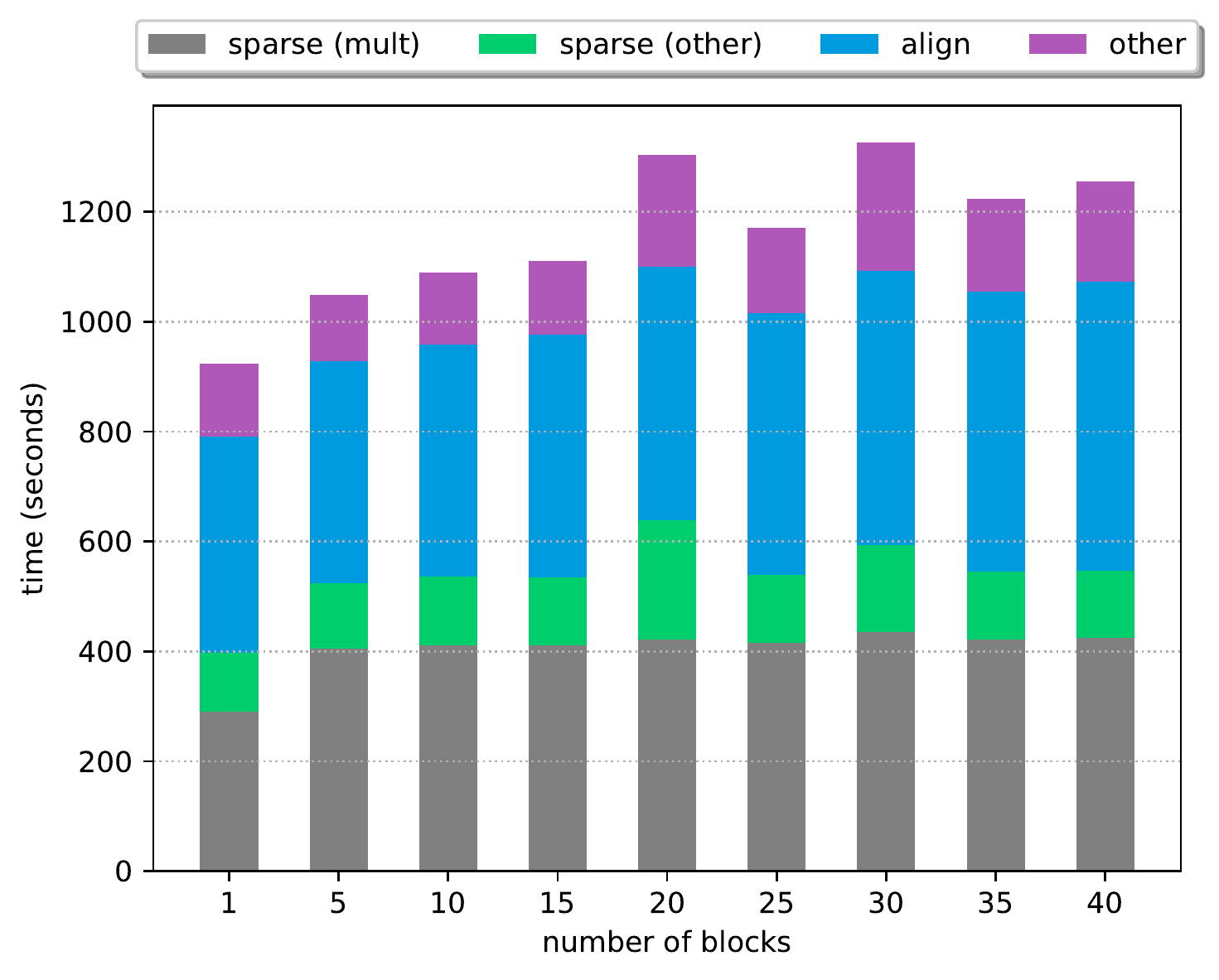}
	\caption{The effect of increasing number of blocks on the runtime of sparse and alignment components.}
	\label{fig:inc-block}
	\vspace{-1.5em}
\end{figure}

Figure~\ref{fig:inc-block} plots the parallel runtime of various components of the similarity search against increasing number of blocks on a dataset containing 20 million sequences over 100 nodes of Summit. 
%
%
Compared to performing the entire search at once, i.e., when number of blocks is 1, there is an average increase of 10-15\% and 40-45\% in the runtime of alignment and multiplication, respectively, with the increase in overall runtime being around 30\%.
We note that this search could not be performed on fewer nodes using only one block, which indicates the severity of the  memory required.
%

\subsection{Load balancing techniques}
\label{sec:lb}
The overlap matrix $C$ in the similarity search is computed to generate the candidate sequences that will be aligned.
The rows and the columns of this sparse matrix represent the sequences and each nonzero element of it corresponds to a pairwise alignment that needs to be performed.
The nonzero elements contain custom information that are needed by the alignment or filtering (such as seed locations in the sequences, common $k$-mer counts, etc).
When computed in parallel using the Blocked 2D Sparse SUMMA algorithm, each block of the overlap matrix is distributed among all $p$ processes in the $\sqrt{p} \times \sqrt{p}$ grid.

The overlap matrix is symmetric: the nonzeros $C_{ij}$ and $C_{ji}$ indicate that an alignment needs to be performed between sequences $i$ and $j$.
This has computational implications for how the search is performed in our work.
First, roughly half of the elements in this matrix may not need to be computed in the multiplication.
Secondly, these also need not be aligned.
Finally, with blocked formation of this matrix, good load balancing necessitates custom methods that take into account all the mentioned conditions.
To this end, we propose two different schemes.

\begin{figure}
	\centering
	\includegraphics[width=0.33\textwidth]{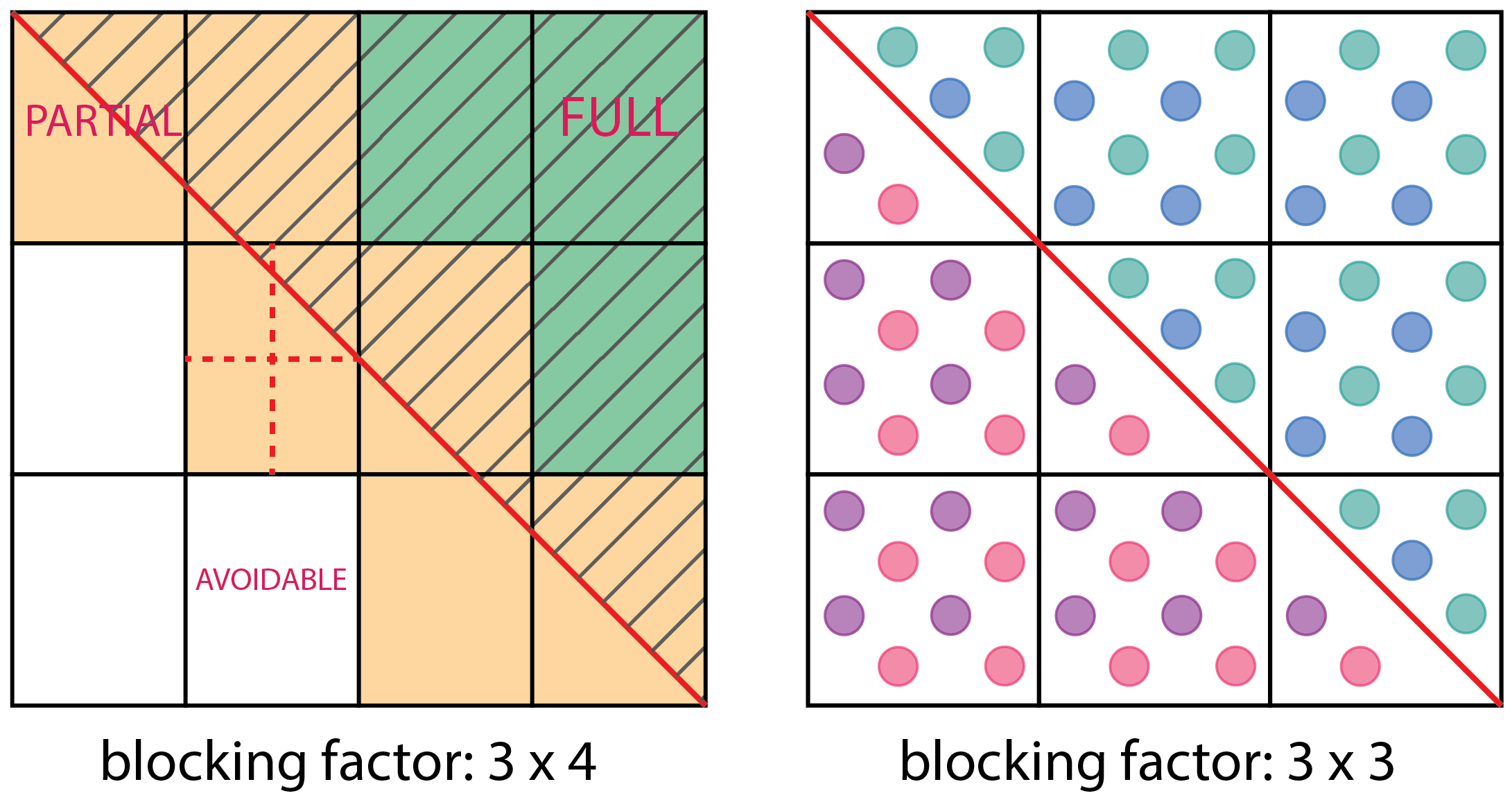}
	\caption{Triangularity-based (left) vs. index-based load balancing (right).}
	\label{fig:lb}
	\vspace{-1.5em}
\end{figure}

\noindent \textbf{Triangularity-based load balancing.} 
In the first load balancing scheme, we only compute the blocks whose intersection with the strictly upper triangular portion of the overlap matrix is non-empty.
The blocks in this way can be categorized into three as full, partial, and avoidable as illustrated in left matrix in Figure~\ref{fig:lb}, where full blocks are colored in green, partial blocks in yellow, and avoidable blocks in white.
%
%
The full and the partial blocks need to be computed in the Blocked 2D Sparse SUMMA, while the avoidable blocks are neither computed nor aligned.
The elements in the full blocks all require an alignment while the the elements in the partial blocks may or may not require an alignment depending on they are in the lower or upper triangular portion of the overlap matrix.
As these blocks are distributed among all processes in the process grid, the partial blocks may lead to load imbalance especially when their intersection with the strictly upper triangular portion is small.
For instance in the overlap matrix in left of Figure~\ref{fig:lb}, assuming a $2 \times 2$ process grid, three processes will stay idle when performing alignments in the block that is in the intersection of second row and column.
In contrast, the load balance of full blocks is better than the partial blocks and the number of full blocks grows quadratically with increasing number of blocks while the number of partial blocks grow linearly.
%

\begin{figure}
  \begin{subfigure}[b]{0.23\textwidth}
    \includegraphics[clip,width=1\textwidth]{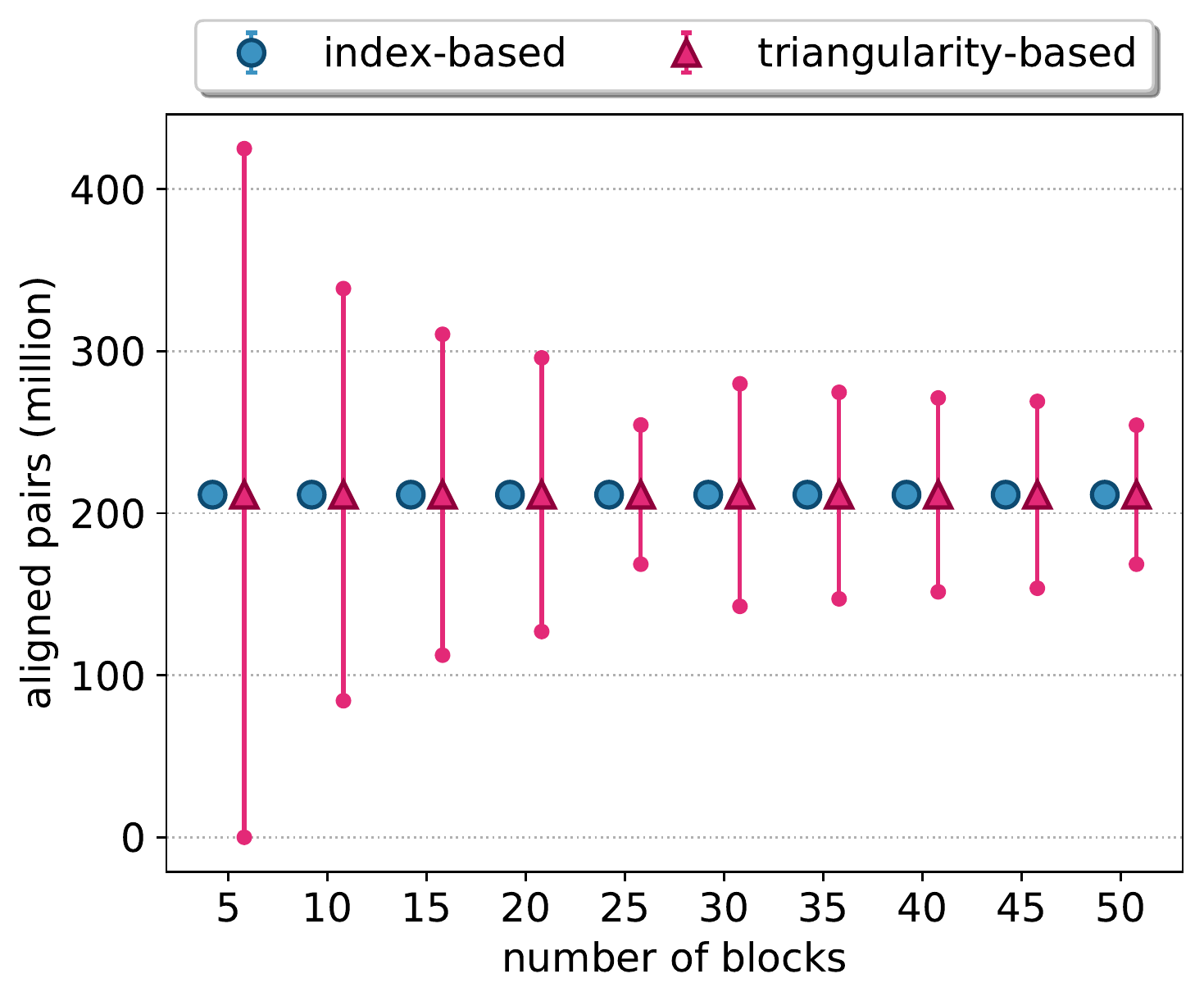}
    \caption{} \label{fig:lb-comp-aln-pairs}
  \end{subfigure}
  \begin{subfigure}[b]{0.23\textwidth}
    \includegraphics[clip,width=1\textwidth]{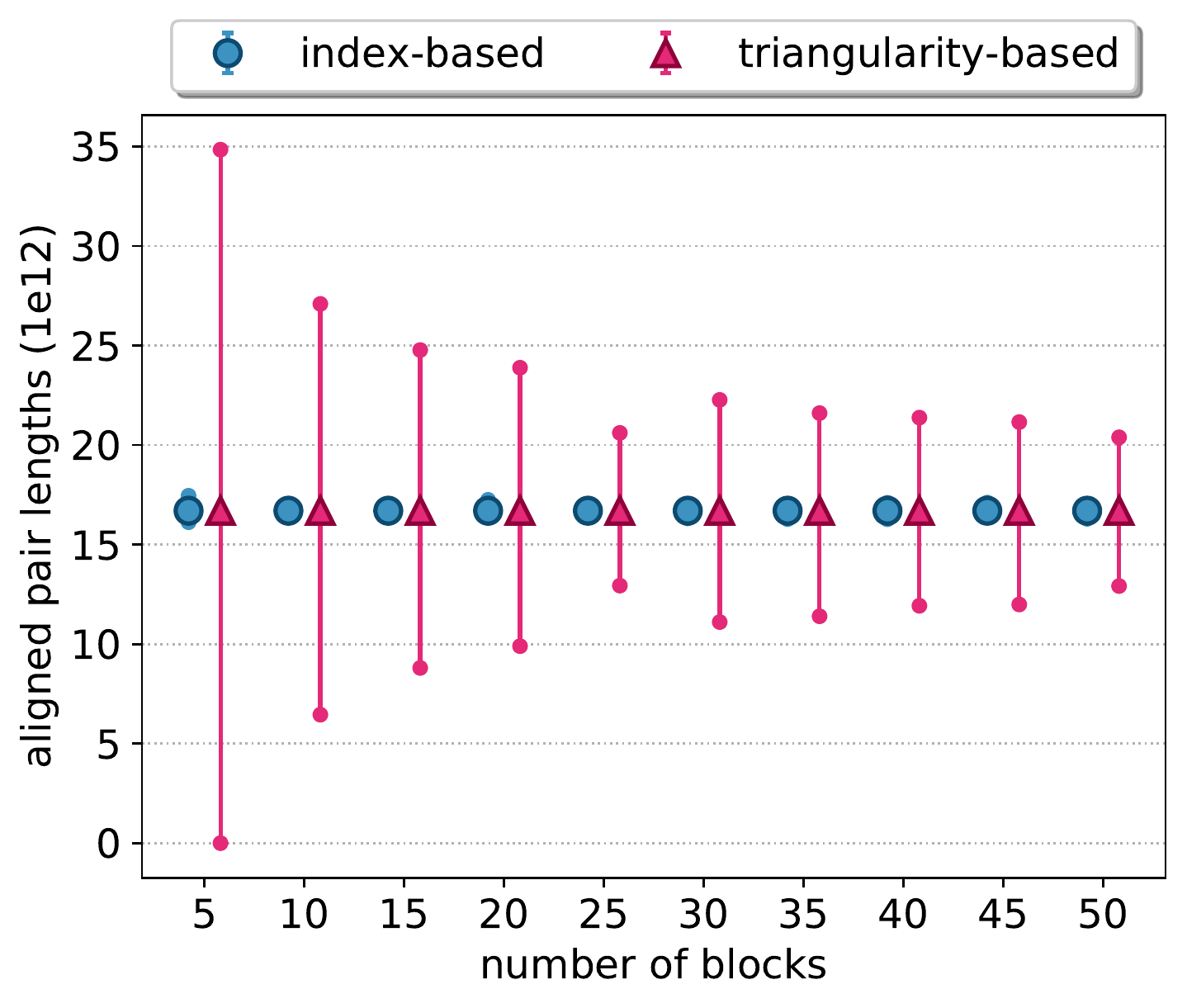}
    \caption{} \label{fig:lb-comp-aln-pair-lens}
  \end{subfigure}
   \begin{subfigure}[b]{0.23\textwidth}
    \includegraphics[clip,width=1\textwidth]{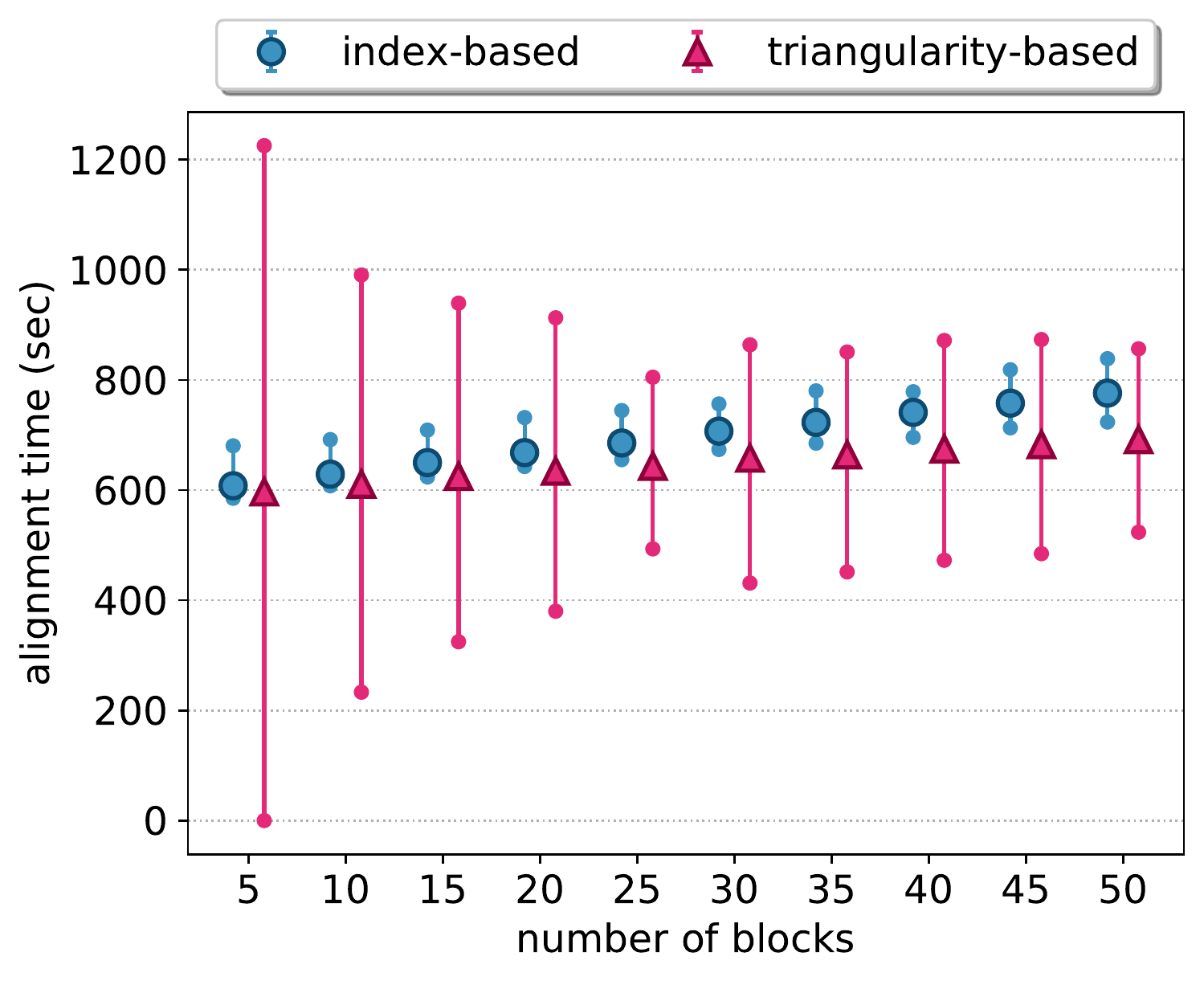}
    \caption{} \label{fig:lb-comp-aln-time}
  \end{subfigure}
   \begin{subfigure}[b]{0.23\textwidth}
    \includegraphics[clip,width=1\textwidth]{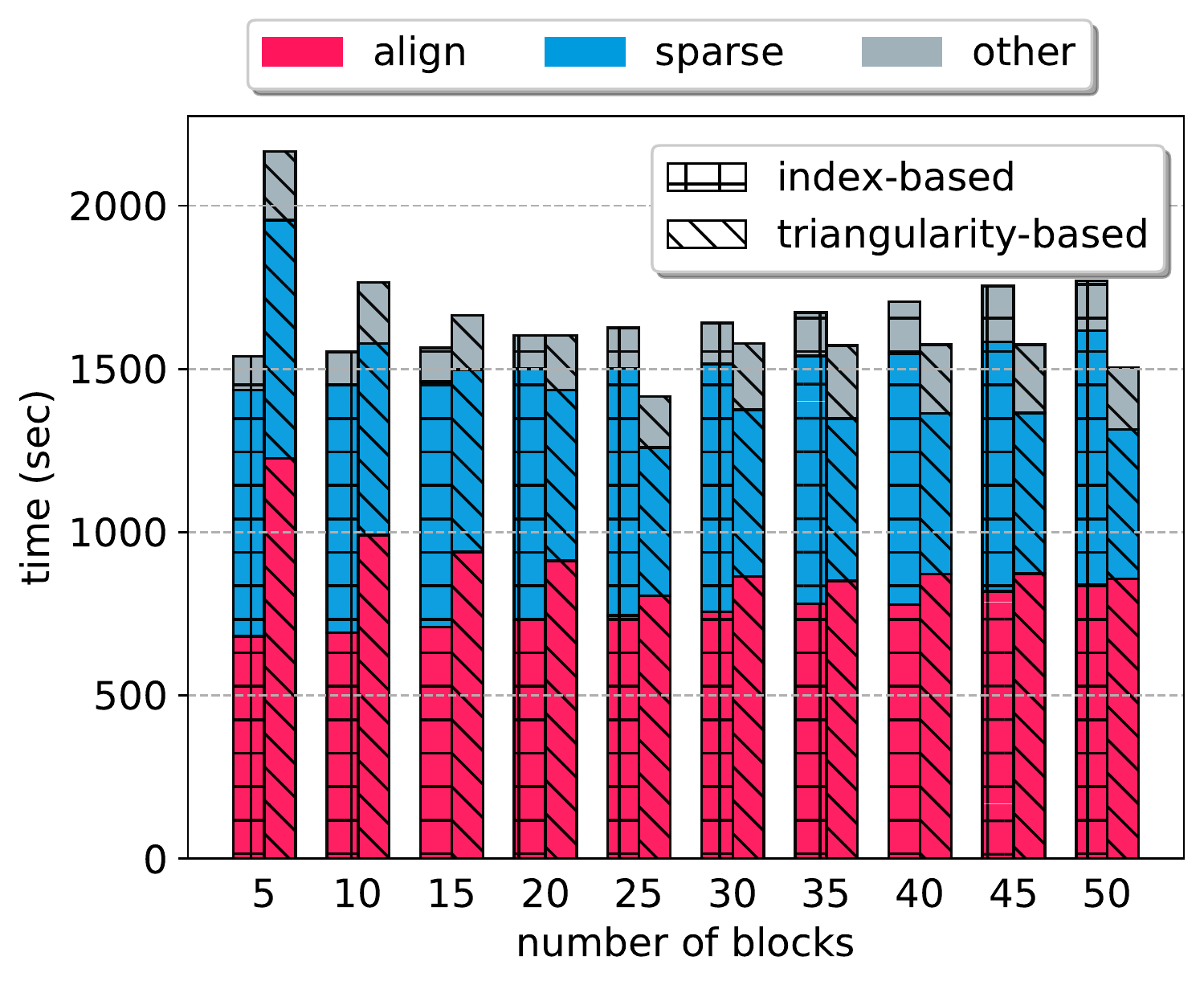}
    \caption{} \label{fig:lb-comp-tot-time}
  \end{subfigure} 
  \caption{Comparison of two load balancing schemes on 64 processes. The three points on a vertical line in the plots at the top and bottom left illustrate the load imbalance by capturing the minimum, average, and maximum values attained by the parallel processes in the respective metric that is measured.
  \label{fig:lb-comp}}
 \vspace{-2em}
\end{figure}

\noindent \textbf{Index-based load balancing.} 
In the second load-balancing scheme, we compute all the blocks and prune the blocks in a  manner to preserve the original nonzero distribution of the overlap matrix, which is usually uniform.
In the lower triangular portion of the matrix, we keep a nonzero if its row and column indices are both odd or both even; and in the upper triangular portion of the matrix, we keep a nonzero if its row index is odd and its column index is even or vice versa.
This process is illustrated at right matrix in Figure~\ref{fig:lb} for a $3 \times 3$ blocking.
This scheme prunes roughly half of each block while respecting the symmetricity of the matrix and ensuring each pair of sequences will be aligned only once.

\noindent \textbf{Comparison}.
The two proposed load-balancing schemes incur same amount of alignment computations.
The triangularity-based load balancing scheme favors saving from sparse computations at the expense of sacrificing from load balance in the partial blocks.
The load balance in full blocks of this scheme however should be as good as the load balance in blocks of the index-based load balancing.
The index-based load balancing aims for better load balancing at the expense of computing each block.
Another aspect these two schemes differ is how they change the structure of the blocks of the overlap matrix.
The index-based scheme preserves the uniform structure of the blocks in the overlap matrix to a large extent, which should lead to better memory access pattern compared to the triangularity-based load balancing.

We compare these two load-balancing schemes on a dataset containing 20 million sequences on 64 nodes of Summit in Figure~\ref{fig:lb-comp}.
%
%
In Figures~\ref{fig:lb-comp-aln-pairs},~\ref{fig:lb-comp-aln-pair-lens}, and \ref{fig:lb-comp-aln-time}, each vertical bar in the plots illustrate the average, minimum, and maximum quantities attained by the parallel processes in the respective metric.
As seen in Figure~\ref{fig:lb-comp-aln-pairs}, the index-based method is able to attain better load balance than the triangularity-based method for all tested block counts. 
The load balance of the triangularity-based method tends to get better with increasing number of blocks.
This is because the ratio of the partial blocks, which are the main cause of load imbalance in this method, decreases with increasing number of blocks.

Although alignment time can be said to be directly proportional to the number of alignments, a better metric is the summation of the sizes of the edit distance matrices as the sequence lengths are different from each other and the alignment algorithm used in this work is a variant of the Smith-Waterman algorithm~\cite{Awan2020} which computes the entire distance matrix.
The load imbalance in this metric is presented in Figure~\ref{fig:lb-comp-aln-pair-lens}.
The index-based method has again better load balance than the triangularity-based method in this metric.
Note that the objective of both load balancing methods is the number of aligned pairs (i.e., Figure~\ref{fig:lb-comp-aln-pairs}) and the index-based method achieves very good load balance in it.
This is reflected in the load imbalance in actual time spent in alignment in Figure~\ref{fig:lb-comp-aln-time}, where the index-based method attains better performance.

\begin{table*}[!tbp]
  \caption{The effect of pre-blocking for index- and triangularity-based load balancing methods.}
  \vspace{-3ex}
  \begin{center}
    \scalebox{0.9} {
      \begin{tabular}{c c r r r r r r r r r r r r}
        \toprule
        & & \multicolumn{4}{c}{time w/o pre-blocking (sec.)} & \multicolumn{4}{c}{time w/ pre-blocking (sec.)} &  \multicolumn{3}{c}{normalized} & \\
          \cmidrule(lr){3-6} \cmidrule(lr){7-10} \cmidrule(lr){11-13}
        load balancing & blocks & align & sparse & sum & total & align & sparse & sum & total & align & sparse & total & efficiency (\%) \\
         \midrule                                                              
         \multirow{5}{*}{index-based} & 10 & 627 & 582 & 1209 & 1555 & 722 & 663 & 740 & 1090 & 1.15 & 1.14 & \textbf{0.70} & 97.6 \\
                                      & 20 & 667 & 582 & 1249 & 1606 & 765 & 726 & 793 & 1123 & 1.15 & 1.25 & \textbf{0.70} & 96.4 \\
                                      & 30 & 705 & 586 & 1291 & 1659 & 804 & 767 & 842 & 1163 & 1.14 & 1.31 & \textbf{0.70} & 95.5 \\
                                      & 40 & 740 & 590 & 1330 & 1724 & 836 & 801 & 873 & 1203 & 1.13 & 1.36 & \textbf{0.70} & 95.7 \\
                                      & 50 & 774 & 596 & 1370 & 1774 & 871 & 841 & 919 & 1245 & 1.13 & 1.41 & \textbf{0.70} & 94.8 \\    
        \midrule
        \multirow{5}{*}{triangularity-based} & 10 & 610 & 465 & 1076 & 1812 & 674 & 610 & 864 & 1468 & 1.10 & 1.31 & \textbf{0.81} & 78.0 \\
                                             & 20 & 634 & 411 & 1045 & 1641 & 694 & 571 & 844 & 1320 & 1.09 & 1.39 & \textbf{0.80} & 82.2 \\
                                             & 30 & 658 & 394 & 1052 & 1602 & 716 & 574 & 857 & 1287 & 1.09 & 1.46 & \textbf{0.80} & 83.5 \\
                                             & 40 & 674 & 388 & 1062 & 1609 & 731 & 585 & 867 & 1286 & 1.08 & 1.51 & \textbf{0.80} & 84.3 \\
                                             & 50 & 692 & 362 & 1053 & 1548 & 749 & 568 & 844 & 1243 & 1.08 & 1.57 & \textbf{0.80} & 88.7 \\

        \bottomrule
        \end{tabular}
    }
  \end{center}
  \label{tb:pb}
  \vspace{-1.5em}
\end{table*}

Finally, the effect of the main advantage of the triangularity-based method, being able to save sparse computations, can be seen in Figure~\ref{fig:lb-comp-tot-time}.
It is able to attain shorter runtime for sparse computations as it is able to avoid a great deal of such computations.
When the effect of this method's load imbalance is low (i.e., when number of blocks is high), it is able attain better total runtime than the index-based method.
As seen in Figure~\ref{fig:lb-comp-tot-time}, for block counts $\{5,10,15,20\}$, the index-based method attains better runtime while in the rest the triangularity-based method attains better runtime due to its ability to avoid sparse computations despite its longer alignment time.

\subsection{Pre-blocking}
\label{sec:pb}
The blocked formation of the similarity graph results in an iterative pipeline in which the graph is constructed incrementally (as seen in Figure~\ref{fig:blocked-summa}).
The components of this pipeline are executed on CPU or GPU resources on the nodes.
The heterogeneous node architecture and the capability to perform the compute-bound batch pairwise alignments--which are amenable for SIMD type of parallelism--on GPUs allow our approach to perform similarity search by utilizing all compute and memory resources on a node.
We further propose an optimization technique, which we refer to as \emph{pre-blocking}, based on the pre-computation of the sparse blocks containing candidate pairs.
The goal of pre-blocking is to increase the efficiency of resource utilization on a node and hence reduce the overall search time.

In the incremental formation of the similarity graph, the alignments are performed on GPUs after discovering them through sparse computations, which are performed on CPUs.
While the alignments are performed in a distributed manner, a big portion of CPU resources is idle and the sparse computations regarding the next block or blocks can actually begin.
In this way, the candidate pairs for the next set of alignments can be discovered in advance and made ready for alignment.
Note that this discovery is a fully-fledged distributed \spgemm with its collective communication operations and memory-bound computations.
Hence, the ability to hide them can prove invaluable by resolving both of these bottlenecks at the same time.

\noindent \textbf{Thread management.}
PASTIS and the graph library used for sparse computations CombBLAS heavily rely on OpenMP for on-node parallelism. 
Although the alignment library ADEPT performs alignments on GPUs, it still uses CPU resources for pre- and post-processing.
Specifically, it uses as many C++ threads as there are GPUs on a node.
We dedicate as many threads as the number of devices on a node to ADEPT and use the rest of the threads for pre-blocking.
The affinities of the threads used by ADEPT are set according to which CPU socket they are attached to.
%

\noindent \textbf{Comparison.}
The main trade-off of the proposed pre-blocking technique is that at the expense of slightly using more memory (the memory required to compute and store the next block or blocks), it hides the \spgemm overhead used in discovering candidate sequence pairs.
Although the similarity search requires extensive amount of memory, the extra memory consumption of the pre-blocking should be low as long as the number of pre-computed blocks is small.
On Summit, we found that in most of our runs the time spent in alignment and sparse computations have a ratio of no more than 2:1.
The number of blocks to pre-compute should depend on this ratio and can be adjusted. 
In our experiments we always pre-compute only the next block.
The pre-blocking is expected to increase the time spent in both the alignment and the sparse computations because these components are now computed concurrently and the computational and memory resources on the CPU have to be shared by both.
However, the overall runtime with the pre-blocking is reduced from the summation of these components to the maximum of them, even if that maximum is slightly increased.

\begin{figure*}
\centering
  \begin{subfigure}[b]{0.27\textwidth}
    \includegraphics[clip,width=1\textwidth]{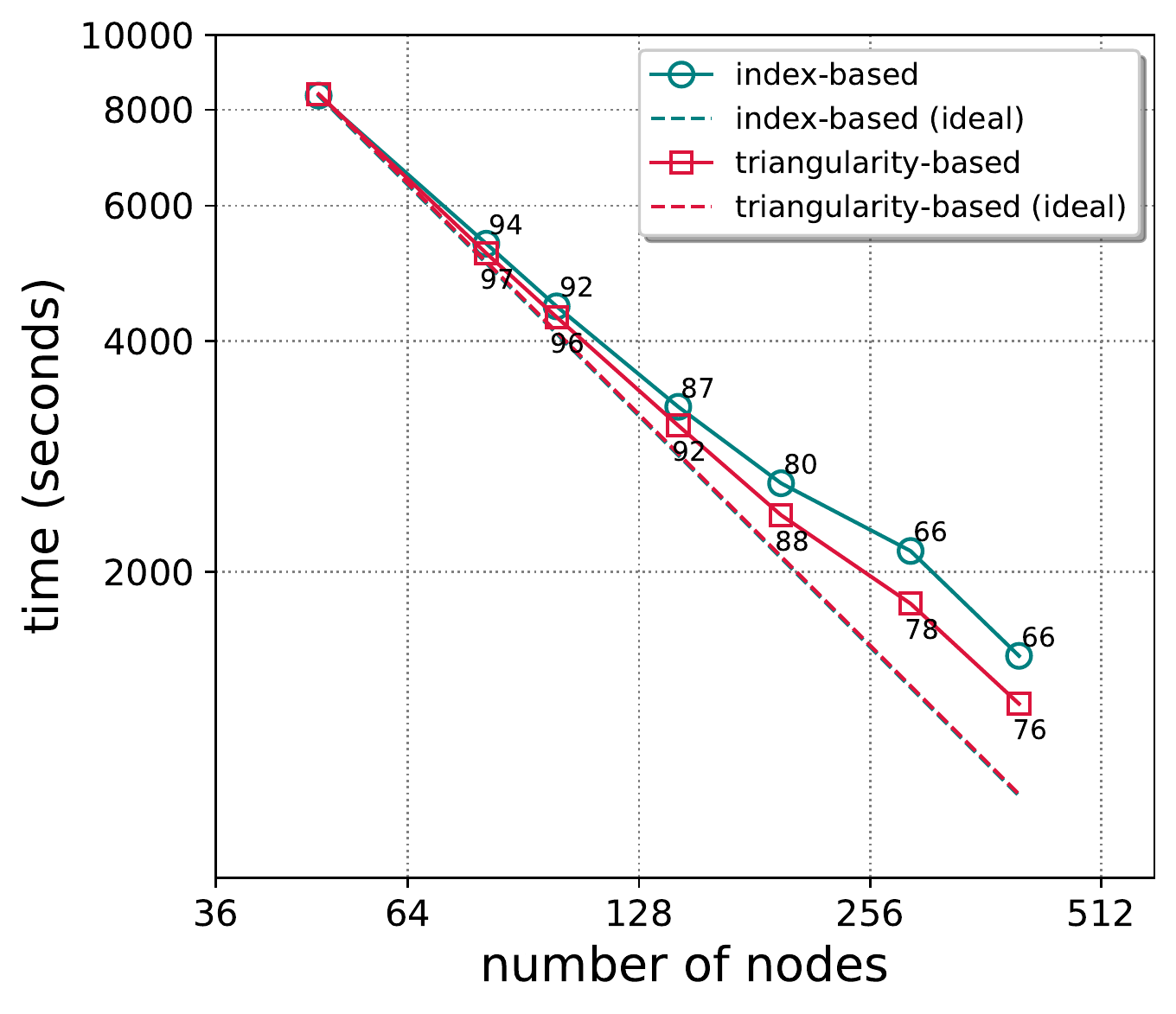}
    \caption{} \label{fig:ss-total}
  \end{subfigure}
      \quad \quad  
  \begin{subfigure}[b]{0.22\textwidth}
    \includegraphics[clip,width=1\textwidth]{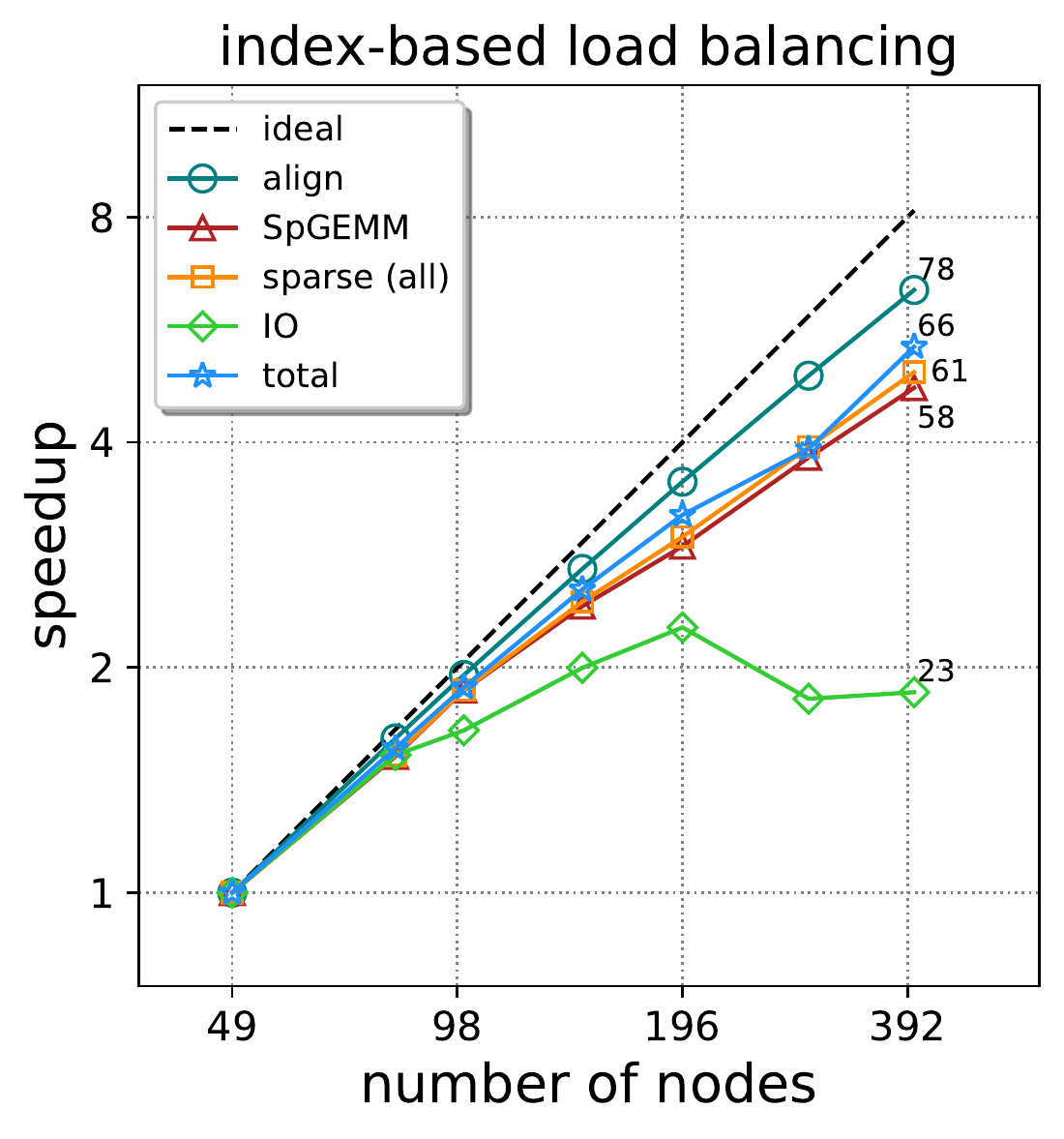}
    \caption{} \label{fig:ss-idx}
  \end{subfigure}
          \quad \quad  
   \begin{subfigure}[b]{0.22\textwidth}
    \includegraphics[clip,width=1\textwidth]{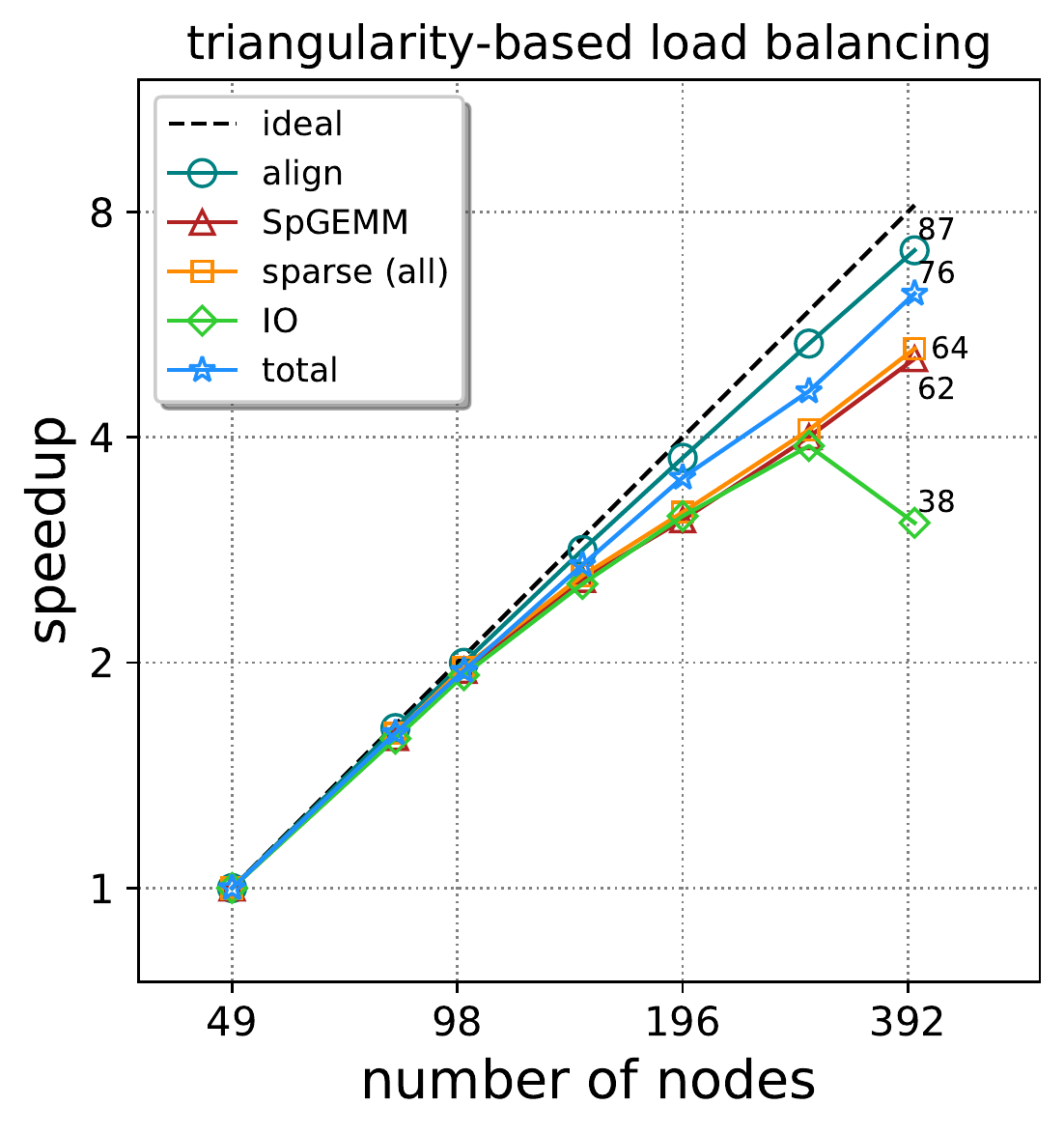}
    \caption{} \label{fig:ss-trg}
  \end{subfigure}
  \caption{Strong scaling performance. The annotated values in the plots indicate the attained parallel efficiency (\%).
  \label{fig:ss}}
 \vspace{-1.5em}
\end{figure*}

Table~\ref{tb:pb} presents various metrics that evaluate the efficiency of the pre-blocking technique.
We compare index- and triangularity-based load balancing methods with and without pre-blocking for five different number of blocks $\{10, 20, 30, 40, 50\}$.
We assess the time spent in alignment, sparse multiplication, summation of these two, and the overall execution time (in the first two big column titles).
Note that the ``sum'' column with the pre-blocking gives the actual obtained time instead of the plain summation of the ``align'' and ``sparse'' columns.
Under the normalized column title, the values obtained with pre-blocking  is normalized with respect to those obtained without pre-blocking.
Finally, the last column evaluates the efficiency of the pre-blocking technique.

As seen in Table~\ref{tb:pb}, although the pre-blocking scheme increases the time spent in alignment and sparse computations, it is able to hide the overheads of the former to a great extent and reduce the runtime by 30\% for the index-based scheme and 20\% for the triangularity-based scheme.
The efficiency of the pre-blocking in the triangularity-based load balancing scheme is lower than that of the index-based scheme (around 80\% vs. 95\%), which can be explained by the fact that the load imbalance found in that scheme adversely affects the efficiency of pre-blocking.
In summary, the pre-blocking technique can be said to be more effective for the index-based load balancing and in both load balancing schemes it is able to reduce the overall runtime significantly.

\section{How Performance Was Measured}
\label{sec:perf-measure}
There are three types of reporting mechanisms used in our work.
The first type is the timers.
These simply measure the elapsed time for certain types of components such as the time spent in alignment, sparse computations, IO, or the total runtime, etc.
The load imbalance for some of the experiments in Section~\ref{sec:innov} is taken by measuring the minimum, average, and maximum time spent in the respective component by all the processes.
The second reported metric is alignments performed per second.
In this metric, we consider the entire parallel runtime and record the number of total pairwise alignments performed and divide the latter by the former.
The final metric is cell updates per second.
This metric is typically utilized within the context of measuring performance of alignment algorithms and it indicates how many cells the utilized algorithm updates in a second.
For this metric we only use the time spent in the alignment kernel, i.e., the forward scoring time in the Smith-Waterman algorithm and divide the number of updated cells by this value.

\section{Performance Results}
\label{sec:perf-res}
This section increases the scale of the experiments conducted compared to the evaluations in Section~\ref{sec:innov} to examine the parallel performance of PASTIS.
We conduct our evaluation on the IBM system Summit at OLCF.
This system consists of 4608 IBM Power System AC922 nodes and each node is equipped with two 22-core 3.8 GHz IBM POWER9 processors  and six NVIDIA Tesla V100 accelerators each of which has 80 streaming multiprocessors.
On each node there is 512 GB of CPU memory and a total of 96 GB of HBM2 memory for accelerators.
The nodes are connected with a dual-rail InfiniBand network in a non-blocking fat tree topology.

We investigate the strong and weak scaling behavior in Section~\ref{sec:ss} and \ref{sec:ws}, respectively.
For these experiments, we stay below 1000 nodes with the largest number of nodes used for strong scaling experiments being 400 and for weak scaling 784.
In the last part of this section (Section~\ref{sec:ls}), we demonstrate our full-scale run using 3364 nodes of Summit.

\subsection{Strong scaling}
\label{sec:ss}
We assess the strong scaling performance for both the index-based and the triangularity-based load balancing schemes.
We use a dataset containing 50 million sequences and scale our approach on $\{49, 81, 100, 144, 196, 289, 400\}$ nodes.
We use 1 MPI task per node and use all the cores and accelerators on each node.
We use a blocking factor of $8 \times 8$ in forming the similarity graph with pre-blocking enabled.
The number of performed alignments for this dataset is 86.5 billion and the entire overlap matrix (i.e., with all blocks) contains 1.99 trillion elements in the index-based scheme and 1.12 trillion elements in the triangularity-based scheme.
%
%
%

Figure~\ref{fig:ss-total} illustrates the strong scaling of PASTIS by plotting the parallel runtime.
The dashed line in the figure indicates the ideal case.
Scaling from 49 nodes to 400 nodes, the index-based load balancing scheme attains a 66\% parallel efficiency and the triangularity-based load balancing scheme attains a 76\% parallel efficiency.
The better efficiency of the latter can be attributed to its avoidance of significant amount of sparse computations, despite having worse load balance.

In Figures~\ref{fig:ss-idx}~and~\ref{fig:ss-trg} we plot the speedup rates of different components.
%
For both of the schemes, it is seen that the computationally intensive component ``align'' (which is performed on accelerators) exhibits better scalability: 78\% and 87\% parallel efficiency for the index-based and triangularity-based schemes, respectively, on 400 nodes.
%
%
The efficiency of sparse operations is around 60\% for both schemes.
With the proposed algorithmic innovations, we are able to overcome the overhead of the sparse computations--whose runtime constitutes a significant portion of the overall runtime (Figure~\ref{fig:lb-comp-tot-time})--to a large extent and lose only 11\%-12\% efficiency due to them.
We note that there are sparse computations that cannot be avoided via pre-blocking.
Although the IO scalability is somewhat erratic, this component constitutes a very minor portion of the overall runtime to be a bottleneck.

\begin{table}[!tbp]
  \caption{Sequence communication wait (cwait) and IO time percentage in overall runtime.}
  \vspace{-2ex}
  \begin{center}
    \scalebox{0.9} {
      \begin{tabular}{c r r r r}
        \toprule
         & \multicolumn{2}{c}{index-based} & \multicolumn{2}{c}{triangularity-based}  \\
          \cmidrule(lr){2-3} \cmidrule(lr){4-5} 
         \#nodes & cwait\% & IO\% & cwait\% & IO\% \\
         \midrule                                                              
        49 & 0.14 & 0.68 & 0.14 & 1.37 \\
        81 &    0.17 & 0.70 & 0.17 & 1.39 \\
        100 & 0.18 & 0.78 & 0.19 & 1.39 \\
        144 & 0.21 & 0.87 & 0.22 & 1.45 \\
        196 & 0.23 & 0.97 & 0.25 & 1.54 \\
        289 & 0.23 & 1.48 & 0.27 & 1.62 \\
        400 & 0.27 & 1.98 & 0.31 & 2.77 \\
        \bottomrule
        \end{tabular}
    }
  \end{center}
  \label{tb:comm-io}
  \vspace{-2em}
\end{table}

Table~\ref{tb:comm-io} presents the overall percentage of IO time and wait time for sequence communication to complete.
PASTIS does not need the sequences until the alignments begin and it uses nonblocking communication for them by starting their transfer right after reading the input sequences.
The waiting time for these communication operations is negligible.
IO also does not constitute a bottleneck, largely due to efficient MPI IO.
The sum of the percentages of these two components is usually less than 3\% of the overall runtime.

\subsection{Weak scaling}
\label{sec:ws}
We examine the weak scaling behavior of our similarity search pipeline for the index-based load balancing scheme.
We vary the number of sequences as we increase the number of nodes $\{25,49,100,196,400,784\}$.
The number of alignments scales quadratically with the number of sequences.
This can also be said to be valid for a majority of the sparse computations as flops in the complexity of the sparse matrix multiplication is proportional to the number of output elements (assuming the compression factor stays the same).
Hence, we use a factor of $\sqrt{x}$ for the number of sequences when we increase the number of processes by a factor of $x$.
As a result, we start with 20 million sequences at 25 nodes and use 28, 40, 56, 80, and 112 million sequences for 49, 100, 196, 400, and 784 nodes, respectively.
Figure~\ref{fig:wsplot} and Table~\ref{tb:wstable} present the obtained results.

The alignment component exhibits better weak scaling efficiency as seen in Figure~\ref{fig:wsplot}.
Overall, all components except IO can said to be exhibiting good weak scaling behavior.
As mentioned before, IO constitutes a very small portion of the parallel runtime and this issue does not seem to be affecting the overall weak scaling efficiency, which stays above 80\%.

\subsection{Similarity search at scale}
\label{sec:ls}
In this section we perform a many-against-many similarity search on a dataset containing 405 million protein sequences.
This dataset was created by clustering and assembling 1.59 billion protein sequence fragments in more than two thousand metagenomic and metatranscriptomic datasets~\cite{Steinegger2018}.
We use a subset of the non-redundant variant in which the subfragments that can be aligned to a longer sequence with 99\% of their residues and a sequence identity of 95\%  are eliminated\footnote{\url{https://metaclust.mmseqs.org/current_release/}}.
Table~\ref{tb:large-run} presents the parameters used in the experiment and the program, and gives details about the obtained results.

\begin{table}[t]
\begin{minipage}[t]{0.43\linewidth}
\centering
\includegraphics[width=53mm]{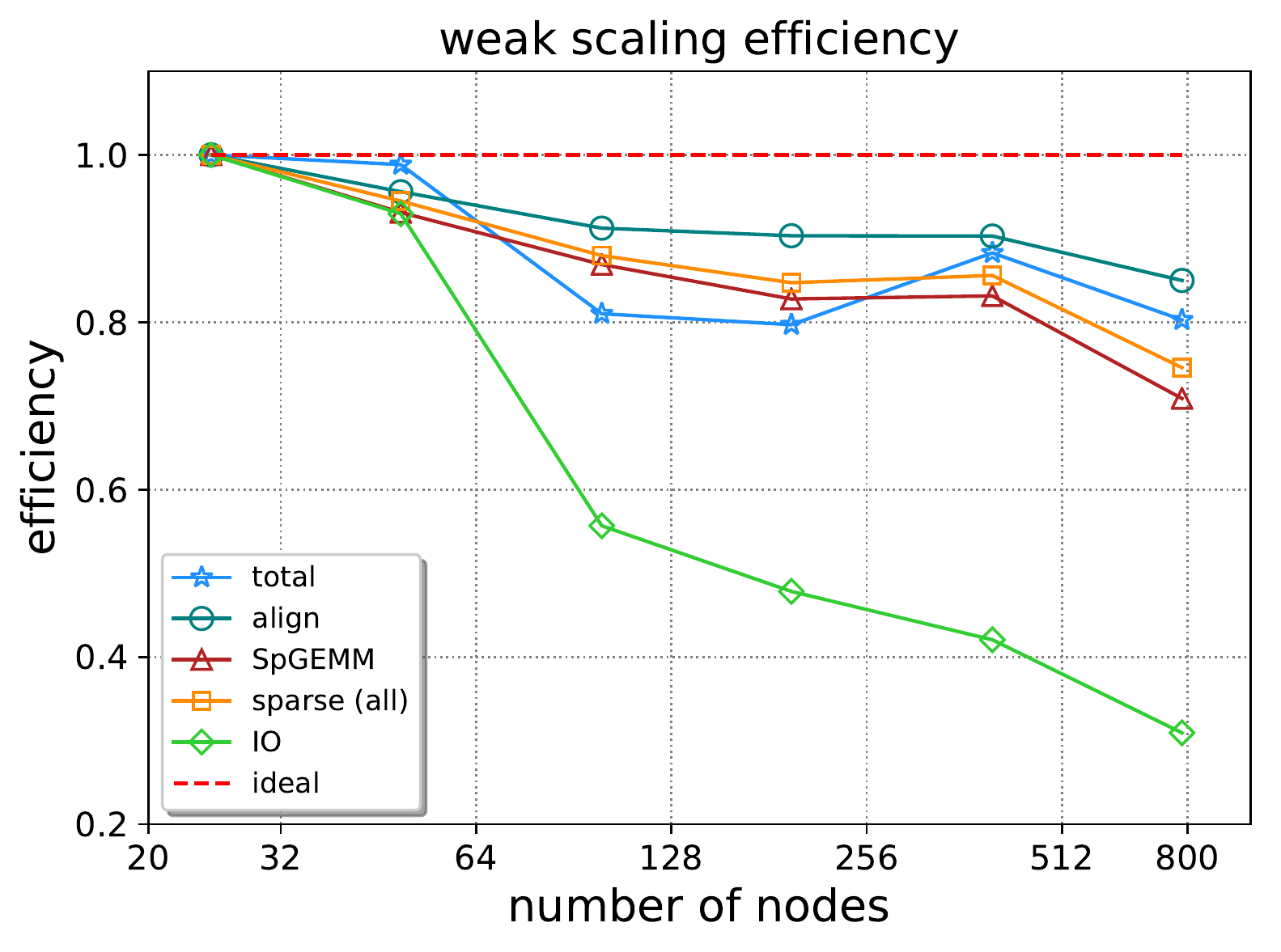}
\captionof{figure}{Weak scaling of different components.}
\label{fig:wsplot}
\end{minipage}
\hfill
\begin{varwidth}[b]{0.37\linewidth}
\centering
\begin{tabular}{ r r r }
    \toprule
    \#nodes & \#seqs. & \#aligns. \\
    \midrule
    25 & 20M & 13.5B \\
    49 & 28M & 26.7B \\
    100 & 40M & 55.1B \\
    196 & 56M & 108.9B \\
    400 & 80M & 225.4B \\
    784 & 112M & 452.4B  \\
    \bottomrule
   \end{tabular}
    \captionsetup{justification=centering}
   \caption{Number of \\ sequences \\ and alignments.}
    \label{tb:wstable}
      \end{varwidth}%
\vspace{-2em}
\end{table}

Our production run used 3364 nodes (73\% of the whole Summit system) and completed the entire search in 3.44 hours.
In total, it discovered 95.9 trillion candidate pairwise alignments, of which it performed 8.6 trillion, and 1.1 trillion of these passed the ANI and coverage thresholds ending up in the final result of the search.
It sustained a rate of 690.6 million alignments per second and achieved a peak rate of 176.3 TCUPs.

We used 1 MPI task per node and at each node 42 cores and 6 GPUs.
We used a total of 400 blocks with a blocking factor of $20 \times 20$ for the Blocked 2D Sparse SUMMA.
For the performance-related parameters that are described in this work, we enabled pre-blocking and utilized triangularity-based load balancing scheme due to its better performance in larger block counts.
A further breakdown of the overall execution time is presented at the bottom of Table~\ref{tb:large-run}.

We attempted to run both DIAMOND and MMSeqs2 on sizable datasets containing 50, 100, and 200 million sequences to perform many-against-many search.
Both of these two search tools rely on SSE and AVX vector instructions for fast alignments.
These instruction sets do not exist on the processors of the Summit system.
For this reason, we tried using Cori system at NERSC, which is a CPU-based system with Intel processors and support for these types of vector instructions.
For MMSeqs2, we started with small number of nodes, i.e., 64 nodes for the 50 million sequence subset but it was not able to complete this run in 6 hours.
We also tried 50 million and 100 million sequence subsets on 256 nodes but again they were not able to complete in 12 hours.
We used a sensitivity value of 5.7 for MMSeqs2 in these tests.
For DIAMOND, we tried 100 million sequence subset on 150 nodes and 200 million sequence subset on 400 nodes with both failing with errors.
We tried both very sensitive and ultra sensitive modes for DIAMOND.
We were able to complete the 50 million sequence subset for DIAMOND on 4 nodes in the default mode (fast mode).
This run completed in 22 minutes sustaining around 60k alignments per second.
This value was much lower than what the authors obtained when running on higher sensitivity modes and larger number of nodes.
For this reason, we compare our run results with those of DIAMOND which are reported very recently on an another supercomputer system Cobra at the Max Planck Society~\cite{Buchfink2021}.

\begin{table}[!tbp]
  \caption{Parameters, results, and statistics of our large-scale production run.}
  \vspace{-2ex}
  \begin{center}
    \scalebox{0.87} {
      \begin{tabular}{l l}
        \toprule
        \multicolumn{2}{c}{\textbf{Experiment parameters}} \\
        \midrule
        System & Summit at OLCF \\
        Number of nodes & 3364 \\
        Process grid (2D) & $58 \times 58$ \\
        Cores per process & 42 \\
        GPUs per process & 6 \\
        Compiler (CPU) & GNU gcc 9.1.0 \\
        Compiler (GPU) & CUDA nvcc 11.0.3 \\
        MPI & Spectrum MPI 10.4 \\
        
        \midrule
        \midrule
        
        \multicolumn{2}{c}{\textbf{Program parameters}} \\
        \midrule
        Number of input sequences & 404,999,880 \\
        $k$-mer length & 6 \\
        Gap open penalty & 11 \\
        Gap extension penalty & 2 \\
        Common $k$-mer threshold & 2 \\
        ANI threshold & 0.30 \\
        Coverage threshold & 0.70 \\
        Blocking factor & $20 \times 20$ \\
        Load balancing & Triangularity-based \\
        Pre-blocking & Enabled \\
        
        \midrule
        \midrule
        
        \multicolumn{2}{c}{\textbf{Results}} \\
        \midrule
        Discovered candidates & 95,855,955,765,012 \\
        Performed alignments & 8,552,623,259,518 (8.9\%) \\
        Similar pairs (output elements) & 1,048,288,620,764 (12.3\%) \\
        Search space & 1.6e17 \\
        Alignment space & 5.2e-5 \\
        Output (file size) & 27 TB \\
        Runtime & 3.44 hours \\
        Alignments per second & 690,609,577 \\
        Cell updates per second & 176.3 TCUPs \\
        
        \midrule
        \midrule
        
        \multicolumn{2}{c}{\textbf{Breakdown \& other}} \\
        \midrule
        \textbf{Time} & \\
        \hspace{0.1cm} Align & 2.62 hours \\
        \hspace{0.1cm} SpGEMM & 2.06 hours \\
        \hspace{0.1cm} Sparse (all) & 2.22 hours \\
        \hspace{0.1cm} Pre-blocking & 2.62 hours \\
        \hspace{0.1cm} IO & 12.0 minutes \\
        \hspace{0.1cm} Communication wait & 0.2 minutes \\
        \textbf{Imbalance (\%)} & \\
        \hspace{0.1cm} Alignment & 7.1 \\
        \hspace{0.1cm} Sparse & 3.1 \\
        \textbf{Sequence by kmer matrix} & \\
        \hspace{0.1cm} Dimensions & 404,988,624 $\times$ 244,140,625 \\
        \hspace{0.1cm} Elements & 48,824,292,733 \\
        
        \bottomrule
        \end{tabular}
    }
  \end{center}
  \label{tb:large-run}
  \vspace{-3em}
\end{table}

We compare the performance results of our run with that of reported by DIAMOND~\cite{Buchfink2021}.
As reported in their work, DIAMOND completed a search of 281 million query sequences against a reference database of 39 million sequences on 520 nodes, taking 5.42 hours and performing 23.0 billion alignments in the very sensitive mode and taking 17.77 hours and performing 23.1 billion alignments in the ultra sensitive mode.
Considering the former, this translates into 1.2 million sequences per second in a many-against-many search space of $281 \times 39 \times 10^{12}$ sequences.
Our run attained 690.6 million sequences per second in a search space of $405 \times 405 \times 10^{12}$ sequences.
Our experiment conducted the search in a space that is 15.0x bigger by increasing the rate of alignments per second by two orders of magnitude, i.e., 575.5x.

As for the total number of alignments performed, if we scale the reported DIAMOND run to the scale of search space our experiment was performed, this becomes equal to a projected value of 345.7 billion alignments.
Our approach can said to perform more sensitive search with alignments per search space being 5.2e-5 compared to DIAMOND's 2.1e-6, amounting to a factor of 24.8x difference.
Time to solution for DIAMOND with an assumption of linear scaling to 2025 nodes results in a projected time to solution of 12.53 hours compared to 3.44 hours of our experiment, which shows our search is 3.6x faster despite performing an order of magnitude more alignments.

\section{Implications}
Many-against-many protein sequence search is a form of sparse and irregular all-vs-all comparisons. The sparsity is data dependent, hence it is known only during runtime which pairwise comparisons will be worth performing. Consequently, it puts a significant stress on the network interconnect. Naively performing this task would amount to a giant \texttt{MPI\_AlltoAllv} call, also known as the personalized all-to-all broadcast. PASTIS significantly reduces this pressure on the interconnect network by regularizing the computation. It does by casting the problem in terms of sparse matrix operations. This technique could also be applied to other irregular applications, as it has been successfully done so for problems in graph and combinatorial problems~\cite{Azad2022}. 

The in-node computation involves sparse matrix-matrix multiplications and a large set of pairwise alignments per node. While the latter maps well to the wide vector units of GPUs, NVIDIA's introduction of Dynamic Programming (DPX) instructions with its newly announced Hopper architecture promises up to 40x speedup for the most expensive part of protein sequence search. PASTIS running on such architectures with DPX instructions would be significantly faster, but also bound by communication costs at scale as the computation speeds up drastically. Hence, future supercomputers that employ accelerators such as Hopper need to provision for higher bisection and network injection bandwidth. 

Since PASTIS uses semiring in SpGEMM, and all the high-performance GPU implementations of SpGEMM being hard-coded for floating-point arithmetic, we performed those steps on the CPU. Thanks to decades of research on high-performance SpGEMM implementations on the CPU, this did not become a performance bottleneck. However, with the aforementioned changes coming to accelerators, sparse matrix operations on other algebras would be a huge welcome. NVIDIA's cuSPARSE library took a big step towards this direction with its latest release where they provide an optimized implementation that performs the multiplication of a sparse matrix and a dense matrix \emph{with custom operators}. We suggest this ability to extend to other functions in the sparse matrix libraries provided by GPU vendors. 

\section*{Acknowledgments}

  This work is supported by the Advanced Scientific Computing Research (ASCR) program within the Office of Science of the DOE under contract number DE-AC02-05CH11231.
  This research was also supported by the Exascale Computing Project (17-SC-20-SC), a collaborative effort of the U.S. Department of Energy Office of Science and the National Nuclear Security Administration. 
  
This research used resources of the Oak Ridge Leadership Computing Facility at
the Oak Ridge National Laboratory, which is supported by the Office of Science
of the U.S. Department of Energy under Contract No. DE-AC05-00OR22725.

\bibliographystyle{IEEEtran}
\bibliography{refs}

\end{document}